\documentclass[preprint,Phys]{SciPost}
\usepackage{graphicx}
\usepackage{dcolumn}
\usepackage{bm,relsize}
\usepackage{amsfonts,amsmath,amssymb,amsthm,mathtools}
\usepackage{slashed}
\usepackage{placeins}
\usepackage{adjustbox}
\usepackage{multirow}
\usepackage{braket}
\usepackage[usenames,dvipsnames,svgnames,table]{xcolor}
\usepackage{breakcites}
\usepackage{appendix}

\hypersetup{
    colorlinks=true,
    linkcolor=ForestGreen,
    filecolor=magenta,
    urlcolor=ForestGreen,
    citecolor=ForestGreen,
}
\usepackage[sort&compress,numbers,merge]{natbib}

\def\beq{\begin{equation}}
\def\eeq{\end{equation}}
\def\bea{\begin{eqnarray}}
\def\eea{\end{eqnarray}}

\begin{document}
%\preprint{APS/123-QED}

\begin{center}{\LARGE \textbf{DarkAgents:\\
\vspace{.1 cm}
\large towards an agentic system for theoretical astroparticle physics}}\end{center}

% TODO: write the author list here. Use initials + surname format.
% Separate subsequent authors by a comma, omit comma at the end of the list.
% Mark the corresponding author with a superscript *.
\renewcommand*{\thefootnote}{\fnsymbol{footnote}}

\begin{center}
\large Michele Lucente,
Silvia Pascoli,
Filippo Sala\footnote{on leave from LPTHE, CNRS \& Sorbonne U., Paris}
,
Matteo Zandi\footnote{matteo.zandi2@studio.unibo.it}
\end{center}

\begin{center}
%{\bf 1}
\textit{Dipartimento di Fisica e Astronomia, Università di Bologna, via Irnerio 46, 40126, Italy\\%\\ and\\
 INFN, Sezione di Bologna, viale Berti Pichat 6/2, 40127, Bologna, 
 Italy}
%* matteo.zandi2@studio.unibo.it
\end{center}

\begin{center}
%\preprintnumber{Prepared for submission to \textit{SciPost Physics}}
\today
\end{center}

\section*{Abstract}
\noindent We present \texttt{DarkAgents}: a multi-agent system that leverages the reasoning and code-generation capabilities of large language models (LLMs), together with deterministic tested human-written code, to build orchestrated pipelines for theoretical astroparticle physics research. While related approaches have been proposed in collider physics and cosmology, \texttt{DarkAgents} targets the specific challenges of this domain, such as model building, complex pipeline computations, multiple constraints and assumption auditing. The framework can be powered by different agentic command-line tools, including Mistral's, Anthropic's, OpenAI's and local LLMs via Ollama. 
%As an initial implementation, we provide \texttt{DarkAgents} with a pipeline receiving as input a classically scale-invariant particle-physics model and providing as output i) the best-fit values of its parameters that reproduce the observed nHz gravitational waves, ii) existing experimental and observational constraints on such parameters, iii) an audit report of the assumptions and priors entering both i) and ii), of particular relevance for astroparticle physics. 
As first implementation, we apply \texttt{DarkAgents} to the study of cosmological first order transitions, starting from a classically scale-invariant particle-physics model and ending with the fit to the NANOGrav nanohertz gravitational-waves spectrum. \texttt{DarkAgent-PT} provides as output i) the best-fit values of model parameters, ii) their existing experimental and observational constraints, iii) an audit report of the assumptions and priors entering both i) and ii), of particular relevance for astroparticle physics.
Our test runs identify inconsistencies in some fits in the literature and produce novel ones based on the dissipative bulk-flow GW template. The code is publicly available at \href{https://github.com/PhysicsZandi/DarkAgents}{https://github.com/PhysicsZandi/DarkAgents}.

%\begin{description}
%\end{description}


%\tableofcontents

\vspace{10pt}
\noindent\rule{\textwidth}{1pt}
\fancyhead[R]{{\bfseries\color{scipostdeepblue}
Prepared for submission to \textit{SciPost Physics}}}
\tableofcontents\thispagestyle{fancy}
%\fancyhead[R]{\nouppercase{\leftmark}}
%\noindent\rule{\textwidth}{1pt}
\vspace{10pt}

\renewcommand*{\thefootnote}{\arabic{footnote}}
\setcounter{footnote}{0}

\clearpage

\fancyhead[R]{\nouppercase{\leftmark}}

\thispagestyle{fancy}

%\tableofcontents

\section{\label{sec:introduction}Introduction}

Theoretical astroparticle physics (TAP) focuses on open questions about the Universe and its fundamental laws and constituents, as well as about its most extreme environments~\cite{AlvesBatista:2021eeu}. In order to look for answers, it brings together knowledge and methods from a very broad spectrum: particle theory and phenomenology, cosmology, astrophysics, gravity and astronomy.
While TAP has reached a very good level of maturity and a deep understanding of the interconnection between its subfields, their co-existence still induces major additional challenges for scientific research compared to a single domain.
%major challenges for scientific research. 

A first challenge is the breadth of the knowledge and methodologies needed for scientific progress, reflected in the proliferation of software tools for very specific aspects of an astroparticle problem, whose connection often needs to be engineered by hand by the researcher. 
A stark contrast is, for example, provided by collider physics phenomenology, where the Beyond-the-Standard-Model (BSM) workflow connects a theoretical model encoded in a Lagrangian with its experimental collider signals, and relies on an extremely well integrated toolchain.

A second challenge is given by the wide range of independent assumptions and uncertainties associated both to observable theoretical predictions and to how we model our Universe.
%how we model the observations of our Universe. 
Examples range from the gravitational-wave templates from phase transitions or topological defects, often applied well beyond their realm of validity, to the fact that the extraction of fundamental-physics knowledge from astrophysical systems often relies on fragile assumptions about them and becomes, therefore, prior-dependent.
This is again very different from i) the quantifiable and controlled uncertainties that accompany theoretical predictions of, e.g., collider and binary GW signals, and ii) with the control of initial states achieved in laboratory experiments, that heavily reduces the prior-dependence of their physics implications.
%\SP{I am not sure what we mean with this paragraph. Contrasting something means that we need then to provide a comparison between the two. Do we mean instead the following?This is very different from i) the quantifiable and controlled uncertainties that accompany theoretical predictions of, e.g., collider and binary GW signals, and ii) with the control of initial states achieved in laboratory experiments, that heavily reduces the prior-dependence of their physics implications. }

These challenges not only hinder newcomers from navigating TAP, but might put under question the validity of some of its findings in broader scientific circles. 
All in all, they ultimately slow down the pace of TAP scientific progress.

\medskip

The revolution of large language models (LLMs), that we are witnessing as we write, opens up a novel realm of possibilities to address the challenges specific to TAP.
Adjacent domains of science have very recently seen the blossoming of end-to-end architectures, that can perform research tasks in theory and phenomenology relying on AI agents, without the need of humans in the loop. These include for example FERMIACC~\cite{Agrawal:2026lvg}, ALBERT\cite{Alexander:2026lpw} and DeepInflation~\cite{Peng:2026ofs} (theory models from measurements at colliders~\cite{Agrawal:2026lvg,Alexander:2026lpw} and in cosmology~\cite{Peng:2026ofs}), ColliderAgent~\cite{Qiu:2026iby} (collider predictions from theory models) and the Mudur et al.~\cite{mudurllm} cosmological framework (predictions from theory and vice-versa).\footnote{
For end-to-end agentic architectures focused on more experimental tasks see CMBagents~\cite{xu2025open} for cosmology and JFC~\cite{Moreno:2026mqk} for colliders (see~\cite{Laverick:2024fyh,Diefenbacher:2025zzn,Gendreau-Distler:2025fsj,Esmail:2026jpb,Hill:2026naa,Badea:2026klb} for previous less complete realizations). 
}

While these efforts are pioneering the path for the partial automation of scientific discovery, to our knowledge none of them addresses the specific TAP challenges identified above.
In order to do so, an end-to-end architecture should
\begin{enumerate}
    \item[a)]
    call deterministic human-written tools in the different domains of TAP, and orchestrate them in a flexible and modular manner that mimics how research is carried out in TAP;
    \item[b)]
    audit the explicit and implicit assumptions in the use of the tools of a) and in the assessment of their physics findings against observations in any of the TAP domains.
    \end{enumerate}

\medskip

In this article we present \texttt{DarkAgents}, the first language-driven multi-agent end-to-end architecture with properties a) and b), that addresses specific questions in TAP going from a theoretical idea to its experimental and observational consequences. 
Here, we focus on a specific physics problem to illustrate our proposal, and present the implementation of \texttt{DarkAgent-PT}, whose input is a particle-physics model and whose outputs are i) the best-fit values of its parameters that reproduce the nHz stochastic background of gravitational-waves (SBGW) observed by NANOGrav in 2023~\cite{NANOGrav:2023gor}, ii) existing experimental and observational constraints on such parameters, iii) an audit report of the assumptions and priors entering both i) and ii). 
The longer-term goal is to develop a DarkAgent community, comprising \texttt{DarkAgent-PT} and many others, each
of which addresses a specific aspect of TAP with an optimised dedicated pipeline. Their communication and interconnection will be orchestrated such that a specific theory idea can be robustly tested against different experimental and observational results.

In addition to the requirements a) and b) demanded by TAP, \texttt{DarkAgents} maintains and facilitates human-control, supervision and direction by being open-access, by providing full written reports and codes for every step of the pipeline, and by being usable with different agentic command-line tools, including Mistral’s \cite{mistral_vibe}, Anthropic’s \cite{anthropic_claude_code}, OpenAI’s \cite{openai_codex} and local LLMs via Ollama \cite{ollama}.  The latter property will also allow us to leverage possible differences in their future development and scaling.

\section{Architecture}

\subsection{Principles and strategy}
Before describing our implementation, we list the criteria that guided its design. The framework needs to be (1) reliable, ensuring correct results and reducing hallucination or silent failure, with all physical quantities produced by deterministic human-validated code; (2) auditable, so that every step can be inspected and reproduced by a human; (3) modular and extensible, allowing new physics pipelines to be added without system re-engineering and letting users run only the steps they need; (4) LLM-agnostic, freeing users from being forced to use a single LLM provider and allowing them to follow the rapidly evolving landscape of available LLM models; (5) efficient and sustainable, enabling a complete analysis to run on a personal computer at the cost of a standard Pro-tier consumer subscription (at the current terms in June 2026).

We identified two possible strategies to build a multi-agent system of this kind. The first is a custom-coded agentic framework, where the orchestration logic, communication between sub-agents and tool calls are all written explicitly as a \texttt{Python} script interfacing with LLMs through API. This gives better control over each sub-agent but requires substantial engineering. The second strategy, which we adopt, uses agentic command-line tools. In this setup, the whole workflow, consisting of the orchestrator, the specialized sub-agents and their skills, is primarily written in plain Markdown instructions and documentation. Agentic command-line tools are also typically equipped with tools for shell access, file reading and writing, web search and fetch, and operate within a local workspace. Each sub-agent is injected with the relevant instructions and skills at the moment it is invoked, while the orchestrator provides the task specification. In collider physics, an example of the former approach is FERMIACC \cite{Agrawal:2026lvg}, whereas an example of the latter is ColliderAgent \cite{Qiu:2026iby}.

\subsection{Orchestrator and sub-agents}
The framework is built around an \texttt{orchestrator}. First, it performs a check of the system
and environment, and installs the packages needed for the deterministic backend to run. Then, it interprets the user's prompt and accordingly selects a supported pipeline branch, writes an explicit execution plan, and invokes the specialized sub-agents in the right order. If it cannot install the required environment, the \texttt{orchestrator} flags this issue in the report and stops, rather than proceeding with an incomplete setup.

The instruction file, named after the sub-agent, contains the scope, the task workflow and rules that the sub-agent needs to follow. Furthermore, the instruction file is provided with information about the upstream inputs that the sub-agent needs to read and the downstream output that it needs to produce: apart from the code files, the sub-agent is asked to produce a human-readable Markdown report for user's inspection and a machine-readable structured JSON file that should follow a fixed scheme, containing the information that the downstream sub-agents will need to read and use. The JSON handoff is checked by the \texttt{orchestrator} after each sub-agent terminates its task. On top of this, the \texttt{orchestrator} pauses the workflow after each sub-agent finishes and provides the user with a summary of the sub-agent's report, giving them the possibility to interact with the sub-agent and ask for clarifications or corrections. It is also possible to render the process autonomous end-to-end by asking the \texttt{orchestrator} not to stop at each step.

%\SP{It is not clear to me what the fixed scheme is} \MZ{A schema is like a dictionary in Python: it is a list in which each object is composed by a name and a value. For example \texttt{{"gauge\_group":"U(1)", "particle\_content":"dark\_scalar", "dark\_photon"}}.} 

The architecture is LLM-agnostic by construction. We give generic names to the orchestrator, the sub-agents and the skills files and to the instruction directory. We map these generic names to the convention of each provider. For example, the orchestrator file is named \texttt{CLAUDE.md} and the instruction directory \texttt{.claude} for Claude Code compatibility, while they are respectively named \texttt{AGENTS.md} and \texttt{.codex} for Codex.

The system is designed to be easily generalizable. Although the current release implements a single pipeline branch, adding or substituting a workflow is straightforward: an implementer should add the set of sub-agents required for the new branch, register the branch with the \texttt{orchestrator} and add compatibility instructions so that the shared upstream and downstream sub-agents can use the new handoffs. 

At this stage, the instructions for each sub-agent are very detailed, so that even less capable AI models can complete the workflow and avoid hallucinations. As future AI models become more capable, the set of instructions can be simplified, while maintaining the deterministic backend and the architecture, keeping \texttt{DarkAgents} on par with AI developments.

%\SP{I think this goes in the discussion of the results and then in the conclusions where we talk about sustainability etc. We rely on the computational speed of the semi-analytic pipeline of
% \cite{Pascoli:2026tuu}, which makes it possible to scan the parameter space and run the inference campaign on a personal computer, without dedicated cluster resources. In practice a single end-to-end run takes of order one hour plus the time of the MCMC campaign, and is feasible with a Pro-tier subscription of any of the supported providers (at the terms current in June 2026).}

\subsection{DarkAgents implementation}

\begin{figure*}[t]
\centering
\includegraphics[width=\linewidth,height=0.35\textheight]{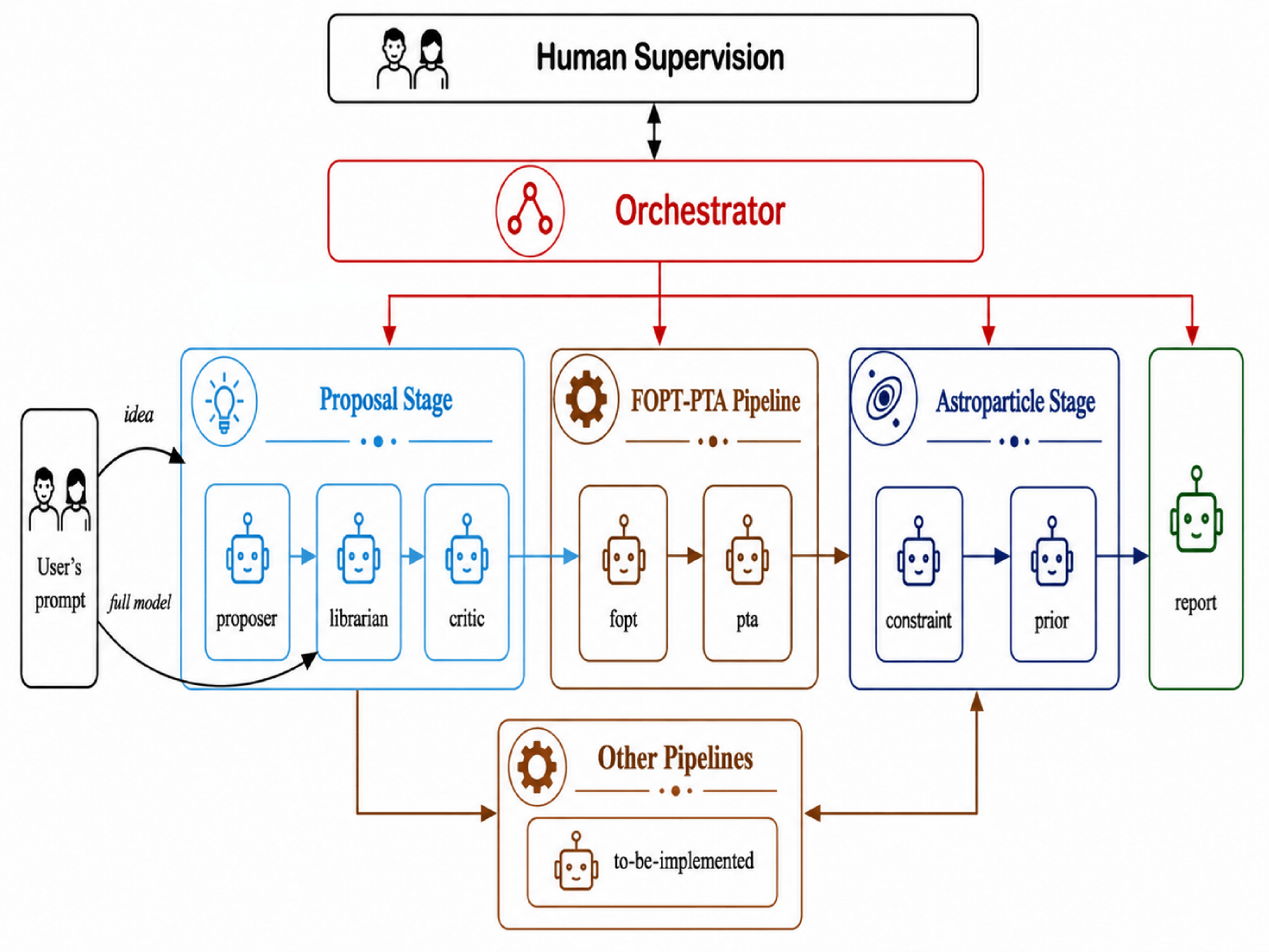}
\caption{Architecture of \texttt{DarkAgents}.
An \texttt{orchestrator} organises and coordinates the workflow of the sub-agents (robot icon in the small rectangular boxes), based on the user's prompts. Humans can audit the progress after each task.
The displayed implementation is for \texttt{DarkAgent-PT}, aimed at studying cosmological FOPT: between the proposal and astroparticle stages, it employs the dedicated FOPT-PTA pipeline. Additional pipelines are shown to illustrate the long-term architectural vision and the expected structure of future versions. 
%\SP{Could we squash the figure slightly vertically? and if so, can we put the humans as an overarching top box?? I also think that the "Othe rpipelines" should fit under the FOPT-PTA box. And I would use the colors a bit differently. I would put the orchestrator in red. Three different colors for the three stages (essentially making Proposal Stage a sligthly different color, and then I would have the pipelines of the same color.}
%\FS{Agree with Silvia's suggestions}
}
\label{fig:architecture}
\end{figure*}

The current version of \texttt{DarkAgents} is depicted in Fig. \ref{fig:architecture}.

The first stage is the proposal one. Users have two options: they can either provide a self-contained well-described particle-physics model or they can give only an idea of the particle-physics model and the target signal they have in mind. In the latter case, a \texttt{proposal-sub-agent} is triggered to identify all the particle-physics information needed for the workflow. In both cases, first a \texttt{librarian-sub-agent} performs a light review of the literature trying to address the novelty of the particle-physics model and whether there are any regions of the parameter space of particular interest. This step helps downstream sub-agents to have an understanding of the current state of the art and gives a hint on where they should start to search in the parameter space, if similar results are present in the literature. After that, a \texttt{critic-sub-agent} tries to find potential issues in the particle-physics model, including QFT inconsistency, and proposes a solution to each of them.

The second stage is branch-dependent. In the current release, only a single branch is present and it implements the pipeline that connects a particle-physics model to its gravitational wave signature due to a cosmological first-order phase transition (FOPT) in classically scale-invariant scenarios, see later for details.

The third stage is the astroparticle one, which is a novel aspect compared to previous implementations in e.g. collider physics. A \texttt{constraint-sub-agent} takes the preferred region of the parameter space and confronts it with current constraints from particle physics experiments, astrophysics and cosmology. It also highlights the possible need for further analysis for a reliable application of such constraints, if the ones available in the literature cannot be directly applied to the given particle model. This step will be further investigated in future versions of \texttt{DarkAgents}: if the \texttt{constraint-sub-agent} flags a missing analysis for which a dedicated branch has been developed, it should communicate with the \texttt{orchestrator} that will interface with the required set of sub-agents to run the workflow of that branch and provide the results as input for another analysis of the \texttt{constraint-sub-agent}. After this part, a \texttt{prior-sub-agent} is invoked to find and audit any assumption, prior, approximation, uncertainty, validity domain, caveat, limitation or warning across the whole workflow that could affect the validity of the final result. Finally, a \texttt{report-sub-agent} writes down a LaTeX report with a summary of the whole workflow.

\subsection{DarkAgent-PT}

As a specific example of the use of \texttt{DarkAgents}, we focus on a cosmological FOPT. We consider the evolution of a scalar field that spontaneously breaks a gauge symmetry in the Early Universe via a FOPT, generating a SGWB. This explanation has been invoked as a possible origin of the nanohertz SGWB observed by NANOGrav \cite{NANOGrav:2023gor} and other PTA experiments in 2023  and subsequently in 2025 \cite{EPTA:2023fyk,Xu:2023wog,Reardon:2023gzh}. The workflow from a theoretical model to its GW signal requires i) to compute the effective potential of the scalar field comprising of loop and finite-temperature contributions; ii) to follow the evolution of the FOPT by tracking the minima of the potential and by computing the bounce action, which controls the nucleation rate due to quantum tunneling or thermal fluctuations; iii) to obtain the relevant cosmological parameters that control the FOPT: transition temperature $T_*$, strength $\alpha$, inverse duration $\beta /H$, the bubble wall velocity $v_w$; iv) to use a template to translate them into a GW spectrum and v) to perform a fit to the data to extract constraints on the particle-physics model parameters. See \cite{Pascoli:2026tuu,Costa:2025csj,Balan:2025uke,Goncalves:2025uwh} for phenomenological studies and \cite{Athron:2023xlk} for a review. 

The \texttt{FOPT-PTA} branch has the goal to apply the described pipeline interfacing with deterministic human-written code. It comprises two sub-agents and their tasks are inspired by the actual work of a theoretical astroparticle physicist. Both sub-agents read the report of the \texttt{librarian-sub-agent}, which allows them to use previous analyses of similar models as initial hints.

The \texttt{fopt-sub-agent} uses the code developed for \cite{Pascoli:2026tuu} to compute the FOPT parameters from the particle-physics model parameters. The semi-analytic fast and efficient implementation in \cite{Pascoli:2026tuu} is crucial at this stage for a very efficient and fast survey of the parameters space without the need of external clusters. Another optimization is given by the fact that \texttt{fopt-sub-agent} provides a preliminary estimate of the expected values of the parameters, by performing a dimensional analysis of the peak frequency of the GW spectrum. 

Then, the \texttt{pta-sub-agent} selects the correct GW spectrum template, based on the acquired knowledge about the FOPT parameters, and it uses it to carry out a first parameter estimate by fitting the spectra within the NANOGrav violins. Once it has identified the correct prior distribution for the particle-physics model parameters, it proceeds to write a dedicated code to use \texttt{PTArcade} \cite{Mitridate:2023oar,Lamb:2023jls} and runs the MCMC scan. After collecting the Bayesian analysis results, it can propose a further scan with an enlarged prior distribution if the $3\sigma$ region touches the boundary of the prior distribution, signaling a too narrow initial choice. After this check, it is able to plot the GW spectrum as a function of the frequency together with the NANOGrav violins for comparison. 

%Both sub-agents read the report of the \texttt{librarian-sub-agent}, which allows them to use previous analyses of similar models as initial hints of the expected allowed region of the parameter space.  
%\SP{ I think this should go earlier , after "It comprises two sub-agents .... physicist." as by giving it here it is a bit confusing.}
%\SP{NEED TO REWRITE THIS. Both the sub-agents are guided by literature findings, but they are instructed to auto-correct if they find a misleading region of the parameter space.} 

For other TAP problems, other branches can be easily implemented as complementary or in substitution to this workflow.

%It is also easier to keep up with the times. While right now the instructions are precise to fulfill a task, it will be much straightforward to relax some instructions and adapt the system to whatever current LLM performs better.  \SP{Can we make this a bit more concrete and detailed?}

%We also suggest sub-agents to perform literature search before and after the pipeline. In this way, sub-agents do not only rely on memory, but compare their assumptions and results with the literature. In fact, literature injection improves the quality and physical consistency of its proposals and critiques while keeping every assumption traceable to a citation. \SP{Should we discuss here the issue of allucinations in literature surveys? Also with relevant references?}

%\SP{It would be good to raise the issues of speed, credits and sustainability and highlight the importance of having fast and efficient numerical tools. } We exploited the computational speed of the semi-analytical pipeline developed in \cite{Pascoli:2026tuu}, thus we only ran on local personal computers.

%\SP{I wonder if it would be easier to merge these two subsections and i) give a list of desiderata at the beginning, ii) then the figure with its detailed description and explanation of the architecture and iii) then how this architecture satisfies the criteria that we set at the beginning. If we do this, the comparison with the other type of architecture (FermiACC) could come here at the end of the section in a discussion paragraph.}

\section{Results}

In order to validate \texttt{DarkAgents}, we tested whether the system can be trusted to execute deterministic code without fabricating the final results and whether the resulting physical statements were consistent.

\subsection{Validation strategy}

It is  known that the use of LLMs intrinsically introduces a source of stochasticity, and their outputs depend on the choice of the LLM model. Therefore, additional checks are needed that are not present when writing deterministic code. We humans carried out a traditional analysis of the same  TAP problem studied by \texttt{DarkAgent-PT}, using the same particle-physics model and deterministic backend, and compared the human and agentic outputs at the level of posterior distributions in Fig. \ref{fig:comparison_human_ai}.
Additionally, we let \texttt{DarkAgents} repeat the same tasks across several LLM providers, and several times with the same prompt, in the period March-June 2026. State-of-the-art models, such as Claude Code (Opus 4.8) and Codex (ChatGPT 5.5), were able to complete the full workflow almost autonomously. Less capable models, such as Mistral Vibe (mistral-medium-3.5), required stronger human guidance and were less reliable in identifying constraints and implicit assumptions, even when they produced a numerically correct run.

We also tested whether the system fails safely: when provided with a model incompatible with the deterministic backend, the orchestrator did not force the task into an incorrect branch, but instead stopped and reported the incompatibility. 

Finally, we tested the literature-injection importance of the sub-agents: when they were supplied with relevant previous analyses and benchmark points, they were inclined to perform their analysis correctly, since they start from a region of the parameter space already surveyed. 

However, we found that using the web tools to search the literature, LLMs were spotted to hallucinate some papers in the reference section of the final report. This indicates that a more thorough examination is needed. 

%\SP{ For me this would go into the results part. Or is it validation????}

\subsection{Example runs} 
%\SP{ I would think we should have a discussion of the results which follows the structure of the architecture step by step. At the moment it seems to me that it is a bit scattered.}

We now report some representative runs of \texttt{DarkAgent-PT}. The purpose of these examples is not to show a complete phenomenological study, but to test if the agentic system can correctly reproduce the \texttt{fopt-branch} by translating a model into backend-compatible code, choosing the appropriate priors, running the inference analysis, identifying the relevant constraints and assumptions. 

Since the initial user's prompt can be given in two different ways, we investigate both possibilities: i) as an explicit particle-physics model, we choose a minimal classically scale-invariant dark $U(1)_D$ model, with the radiatively-generated vacuum expectation value $v$ and the gauge coupling $g_D$ as free parameters; ii) as a generic prompt without specifying a given Lagrangian (referred to as ``idea" in Fig.~\ref{fig:architecture}), we asked to find if a classically scale-invariant $U(1)$ model with a dark scalar, a dark gauge boson and a dark fermion can explain the NANOGrav signal through a cosmological FOPT. We have tried different prompts for ``ideas'', such as non-abelian models, and also in these cases \texttt{DarkAgent-PT} has produced new meaningful results, see reports in the public repository \href{https://github.com/PhysicsZandi/DarkAgents}{https://github.com/PhysicsZandi/DarkAgents}.

\begin{figure*}[t]
\centering
\includegraphics[width=0.49\linewidth]{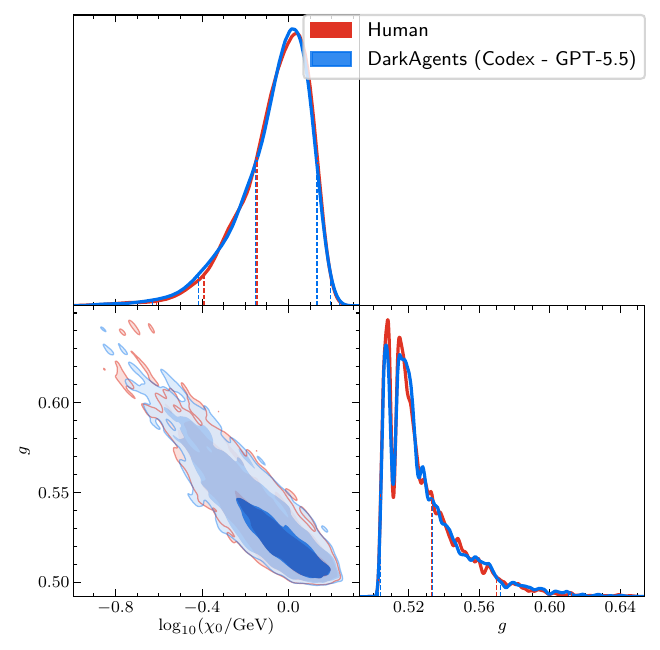}
\includegraphics[width=0.49\linewidth]{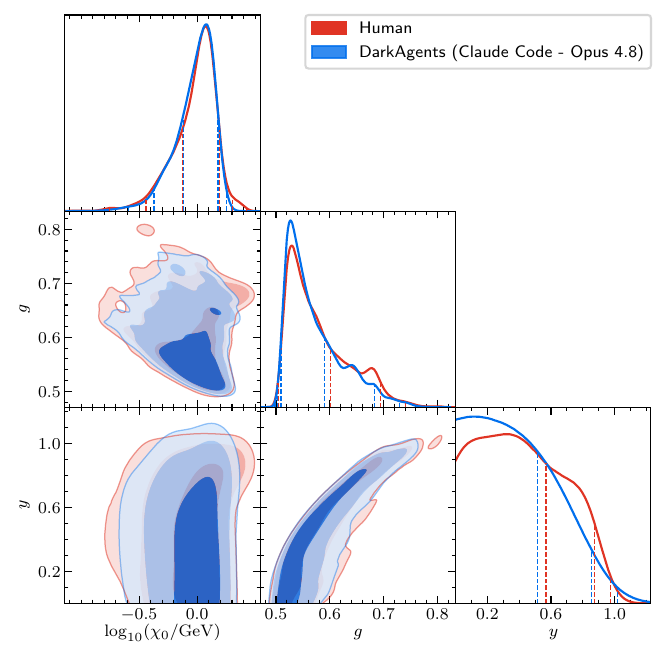}
\caption{Comparison of the Bayesian posterior distributions against the NANOGrav 15yr dataset~\cite{NANOGrav:2023gor} obtained with \texttt{PTArcade}~\cite{Mitridate:2023oar,Lamb:2023jls}, using the (dissipative) bulk flow template~\cite{Lewicki:2025hxg}. On the left the user-provided minimal $U(1)$ model with a dark scalar and a dark photon, whereas on the right the idea-provided $U(1)$ model with the addition of a dark fermion. The red region is computed by a human researcher, whereas the blue one is one of the outputs of \texttt{DarkAgent-PT}.
The online repository \href{https://github.com/PhysicsZandi/DarkAgents}{https://github.com/PhysicsZandi/DarkAgents} contains more fit results, both for more models and including the SMBHB astrophysical contribution of super-massive black-hole binaries.}

%\SP{Could we have log-scale ticks on the x-axis?} \SP{What are the vertical dashed lines?} \SP{I do nor undertsand why for the "idea" one the values of the vev are more constrained. Should it be wider than the minimal U(1) since you could always set the yukawa coupling to zero and recover the minimal model? Or is it a problem of the scanning of the small v region?? \ML{Metterei il modello esatto utilizzato (GPT 5.5?, Opus 4.8?) invece di Codex e Claude Code, che sono i tool su cui girano.}
\label{fig:comparison_human_ai}
\end{figure*}

The comparison between the Bayesian posterior distribution obtained by \texttt{DarkAgent-PT} and by a human researcher for both particle-physics models is reported in Fig. \ref{fig:comparison_human_ai}. 
%\SP{I think we need to comment on the results here.} 
We see that \texttt{DarkAgent-PT} is able to reproduce very closely the traditional research results with very similar Bayesian posterior distributions. 
%\SP{Should we discuss some of the physics results? What about the point of the dissipative bulk flow in defining the range of the vev, w.r.t. literature???}
Beyond generating the correct codes and obtaining right results, the state-of-the-art models demonstrated the ability to make appropriate physical choices. They rejected the sound-wave spectrum template~\cite{Jinno:2022mie,Caprini:2024hue} in the regime of $\alpha$ where it is not valid and selected the (dissipative) bulk-flow one \cite{Lewicki:2025hxg}. They explicitly acknowledged the residual uncertainty in the template choice without carrying out a detailed study. It turns out that such rejection is indeed correct, see Sec. 4 of~\cite{Musumeci:2026jum} for a more detailed discussion. 
%as a traditional physics analysis of bubble dynamics confirms with an in-depth study of the physics~\cite{InPreparation}. 
We further asked \texttt{DarkAgent-PT} to include super-massive-black-hole binaries (SMBHB) in the fit, and in doing so they satisfactorily reproduced traditional results by the NANOGrav collaboration~\cite{NANOGrav:2023hvm} in terms of $\alpha$, $\beta/H$, $T_\text{reh}$ and extended them by i) linking those quantities to $g_D$ and $v$ and fitting directly in terms of the latters ii) by using a more appropriate GW template. 
They critically discussed the impact on its results of the astrophysical assumptions and uncertainties in the GW spectrum from SMBHB, although they needed an additional prompt to do so.

Furthermore, in searching for constraints, we highlight that they were able to identify cosmological constraints due to the relativistic degrees of freedom, $\Delta N_{\text{eff}}$, to reheating before Big Bang Nucleosynthesis and to dark-sector thermal history as well as astrophysical bounds from Supernova (SN1987A) cooling and experimental ones from beam‑dump, fixed‑target, collider experiments  \cite{Agrawal:2021dbo,Antel:2023hkf}.
However, the kinetic and scalar mixing, potentially present in this type of rich dark sector extensions of the Standard Model \cite{Abdullahi:2025fiy}, are not included at this stage and no backend or instruction is given to compute the physical quantities relevant to apply these bounds, such as widths and lifetimes.
The system did not proceed with hallucinations about the impact of these bounds, but raised warnings that there is not enough information to evaluate them and went even further, by indicating which phenomenological analyses need to be carried out in order to assess the impact of those limits.
As discussed already, in the future, additional \texttt{DarkAgents} should be developed and would be activated to provide the detailed computations needed to assess and apply each type of constraint. 
 
Regarding the implicit assumptions, \texttt{DarkAgents-PT} also diagnosed limitations of the (deliberately simplified) version of the pipeline of~\cite{Pascoli:2026tuu} that we provided, correctly noting that a better renormalization-scale choice, resummation of daisy terms, running of the couplings and treatment of gauge dependence should be included, together with the missing bubble wall velocity calculation.

It is non-trivial that \texttt{DarkAgents} has succeeded in obtaining correct results and in meaningfully auditing assumptions both in its pipeline and in the constraints, despite the fact that we designed it with such goals in mind.
While this success admittedly concerns a specific physics problem, it indicates that the architecture we propose may be a step in the right direction of facilitating and accelerating TAP research with the assistance of AI.

A set of complete example runs for both particle-physics models and all LLM providers, including every generated script, report, handoff and figure, is provided in the public repository \href{https://github.com/PhysicsZandi/DarkAgents}{https://github.com/PhysicsZandi/DarkAgents}.

\section{Discussion and outlook}

We have presented \texttt{DarkAgents}: a multi-agent system that leverages new developments in Large Language Models (LLMs) for reasoning and code-generation, to address a given question in theoretical astroparticle physics (TAP).
It is an end-to-end architecture that takes as input a particle physics idea or model, and produces as output a report presenting its phenomenological analysis. It is organized in three stages: a proposal, a pipeline dedicated to a specific problem of interest, and an astroparticle stage, see Fig.~\ref{fig:architecture}.
%\texttt{DarkAgents} mimics how TAP physicists typically work not only in its structure, but also by deciding the best strategy, first with guesses educated by Naive Dimensional Analysis (NDA), and subsequently revising them through an iterative improving procedure, based on the outcome of the previous attempts.
\texttt{DarkAgents} mimics how TAP physicists typically work not only in its structure, but also in methodology: first making guesses educated by Naive Dimensional Analysis (NDA), and subsequently revising them through an iterative improving procedure, based on the outcome of the previous attempts.

Our code is open access and can be found, together with test prompts and reports, at \href{https://github.com/PhysicsZandi/DarkAgents}{https://github.com/PhysicsZandi/DarkAgents}.

\texttt{DarkAgents} is designed to address the specific challenges of TAP that slow down the pace of its progress. One is the breadth of the knowledge and methodologies required for its advancement. This challenge is addressed by calling and orchestrating diverse deterministic human-written tools, as well as by an astroparticle-specific sub-agent, \texttt{constraint-sub-agent}, with the role of identifying and combining a plethora of experimental and observational constraints. Another TAP challenge is the dependence of some of its findings on assumptions. \texttt{DarkAgents} addresses it with a second astroparticle-specific sub-agent, \texttt{prior-sub-agent}, that searches for explicit and hidden physical assumptions in the derivation of the results obtained within the entire architecture, and so identifies the limits of validity of its results.
%The latter is achieved by two astroparticle-specific subagents, constraint and prior. The former has the role of identifying and combining a plethora of experimental and observational constraints, that may or may not apply to the particle physics model under consideration. \texttt{DarkAgents} is able to perform this task efficiently, also highlighting missing analysis required to apply given bounds. The latter searches for hidden astrophysical/cosmological assumptions to assess the validity of the results obtained. 
These sub-agents are able to perform these tasks efficiently, and also to identify the missing phenomenological analysis required to apply existing bounds. 

As an example, we have implemented \texttt{DarkAgents} as \texttt{DarkAgent-PT}, to study scale-invariant scalar extensions of the Standard Model, in which a FOPT at the GeV scale may explain the observation of a nanohertz SGWB by NANOGrav and other PTA experiments. We find that \texttt{DarkAgent-PT} is able to reproduce human-generated results both in the case in which the researcher provides the model, i.e. a $U(1)_D$ minimal model with a complex scalar and a dark photon, and in the case in which \texttt{DarkAgent-PT} is given more freedom to define the model starting from a less structured prompt, see Fig.~\ref{fig:comparison_human_ai}. We find relatively large values of gauge couplings and GeV-scale vevs for the scalar, significantly larger than in analysis that use the sound-wave template.
Remarkably, it was \texttt{DarkAgent-PT} that autonomously and correctly found out that, for those models, the optimal GW template to use is not the sound-wave one often employed in the literature, but the dissipative bulk flow of~\cite{Lewicki:2025hxg}, as pointed out in ~\cite{Musumeci:2026jum}.
While our design succeeds in bringing down silent fails and hallucinations, some drawbacks remain, like made-up references in the final pdf reports.
Ultimate human control and supervision stays essential to guide the developments and to ensure reliability of results.

\medskip

In order to adress the specific challenges of TAP, we designed \texttt{DarkAgents} based on the principles of i) reliability, reducing hallucination
or silent failure, ii) the option of human-audit at every step, iii) modularity and possibility to extend it without system
re-engineering, iv) LLM-agnosticism, that also allows to keep up with rapid developments in the LLM models, and v) efficiency and sustainability. 

 One important driver for the design of \texttt{DarkAgents} has been the lurking question ``When will \texttt{DarkAgents} just be surpassed by the increased power of generic LLM?''.
This question is linked to ``the bitter lesson''~\cite{sutton_bitter_lesson}, that search and learning with little other input have so far proven to become more powerful with increase in computing power, than ad-hoc design with too much engineering. This has pushed us to a design that does not rely on such engineering and that calls at every step an LLM, in order to benefit from their dramatic future scaling.\footnote{For example, it is not inconceivable that eventually the \texttt{prior-sub-agent} will not limit itself to identify inconsistencies between the assumptions of some constraints or tools used, but will go beyond and derive new constraints or propose new tools that are suitable to the problem at hand.}
The bitter lesson has also pushed us further, to make the architecture flexible in different key respects. 
First, it is flexible in the level of human control and supervision. Researchers can audit each step of the pipeline to ensure the reliability of results, but they do not need to if the workflow can be trusted. It is expected that going forward the need of human supervision at intermediate steps will decrease, and \texttt{DarkAgents} has already been designed to adapt to this evolution. Second, the architecture as well as the sub-agents we implemented are flexible in that they work with different agentic tools including Mistral's, Anthropic's, OpenAI's and local LLMs via Ollama. This not only frees the researcher from the need of changing the LLM code but also allows to exploit the most advanced ones at a given moment, keeping in line with the rapid developments in the field.

%The architecture is also flexible in the level of human-control and supervision. Researchers can audit each step of the pipeline to ensure the reliability of results but they do not need to if the workflow can be trusted. It is expected that going forward the need of human supervision at intermediate steps will decrease and \texttt{DarkAgents} has already been designed to adapt to this evolution.

%In addition to this validation of the approach, we highlight some interesting outputs by \texttt{DarkAgents}.
%\SP{Here we should mention the issue of the bulkflow. the many contraints identified. But also the shortcomings in terms of literature hallucinations etc.}

Another driver in shaping our design, especially its modularity, has been our vision that each problem in TAP should have its own \texttt{DarkAgent}. In this sense, \texttt{DarkAgents-PT} is thought of the first one in TAP, by which we want to demonstrate the efficiency and capabilities of the entire architecture for astroparticle studies. 
Together, the \texttt{DarkAgents} would constitute a community that can communicate and interact under an orchestrator, in order to study a given model from all required points of view. To this aim, we advocate for the development of other \texttt{DarkAgents} on specific astroparticle areas and for their Open Access,
%in order to fully exploit the \texttt{DarkAgents} community
in order to fully exploit the extraordinary impact that AI via LLM can have on our ability to perform efficiently and effectively research in astroparticle physics.
We think that such strategy will be of great aid in facilitating and accelerating research in astroparticle physics, in an AI-assisted human-controlled manner.

\medskip

\section*{Acknowledgements}

This work has been partly funded by the European Union under the Horizon Europe’s Project: 101201278 – DarkSHunt - ERC - 2024 ADG and by the Italian INFN program on Theoretical Astroparticle Physics (TAsP). Views and opinions expressed are however those of the author(s) only and do not necessarily reflect those of the European Union or the European Research Council Executive Agency. Neither the European Union nor the granting authority can be held responsible for them. During a significant part of this project, ML was funded by the European Union under the Horizon Europe's Marie Sklodowska-Curie project 101068791 — NuBridge.
FS thanks the public library `Biblioteca delle Oblate' in Firenze for hospitality while part of this work was being completed.

% for the sample file to show the different styles of references, but authors most likely will not want to use it.
%\nocite{*}

%\bibliographystyle{JHEP}
\bibliography{DarkAgents}% Produces the bibliography via BibTeX.

\nolinenumbers
\end{document}